\documentclass[useAMS,preprint2]{aastex}
\usepackage{natbib}

\shorttitle{Discovery of WASP-120b, 122b and 123b}
\shortauthors{O. D. Turner et al.}

\begin{document}
 
\title{WASP-120\,b, WASP-122\,b and WASP-123\,b: Three newly discovered planets from the WASP-South survey}

\author{O.D.Turner\altaffilmark{1},
D. R. Anderson\altaffilmark{1}, 
A. Collier Cameron\altaffilmark{2}, 
L. Delrez\altaffilmark{3},
D. F. Evans\altaffilmark{1} 
M. Gillon\altaffilmark{3}, 
C. Hellier\altaffilmark{1},
E. Jehin\altaffilmark{3},  
M. Lendl\altaffilmark{4,5},  
P. F. L. Maxted\altaffilmark{1}, 
F. Pepe\altaffilmark{5},   
D. Pollacco\altaffilmark{6}, 
D. Queloz\altaffilmark{7},
D. S\'{e}gransan\altaffilmark{5}, 
B. Smalley\altaffilmark{1}, 
A. M. S. Smith\altaffilmark{8,9}, 
A. H. M. J. Triaud\altaffilmark{5,10,11}, 
S. Udry\altaffilmark{5}, 
R. G. West\altaffilmark{6}}

\altaffiltext{1}{Astrophysics Group, Keele University, Staffordshire ST5 5BG, UK}
\altaffiltext{2}{SUPA, School of Physics and Astronomy, University of St. Andrews, North Haugh, Fife KY16 9SS, UK}
\altaffiltext{3}{Institut d'Astrophysique et de G\'{e}ophysique, Universit\`{e} de Li\'{e}ge, All\`{e}e du 6 Ao\^{u}t, 17, Bat. B5C, Li\`{e}ge 1, Belgium}
\altaffiltext{4}{Austrian Academy of Sciences, Space Research Institute, Schmiedlstraße 6, 8042 Graz, Austria}
\altaffiltext{5}{Observatoire de Gen\`{e}ve, Universit\'{e} de Gen\`{e}ve, 51 Chemin des Maillettes, 1290 Sauverny, Switzerland} 
\altaffiltext{6}{Department of Physics, University of Warwick, Coventry CV4 7AL, UK} 
\altaffiltext{7}{Cavendish Laboratory, J J Thomson Avenue, Cambridge, CB3 0HE, UK} 
\altaffiltext{8}{N. Copernicus Astronomical Centre, Polish Academy of Sciences, Bartycka 18, 00-716, Warsaw, Poland}
\altaffiltext{9}{Institute of Planetary Research, German Aerospace Center, Rutherfordstrasse 2, D-12489 Berlin, Germany}
\altaffiltext{10}{Centre for Planetary Sciences, University of Toronto at Scarborough, Toronto, Ontario M1C 1A4, Canada}
\altaffiltext{11}{Department of Astronomy \& Astrophysics, University of Toronto, Toronto, ON M5S 3H4, Canada}

\begin{abstract}
We present the discovery by the WASP-South survey of three planets transiting moderately bright stars (V $\approx 11$). WASP-120\,b is a massive ($4.85 M_{\rm Jup}$) planet in a 3.6-day orbit that we find likely to be eccentric ($e = 0.059^{+0.025}_{-0.018}$) around an F5 star. WASP-122\,b is a hot-Jupiter ($1.28 M_{\rm Jup}$, $1.74 R_{\rm Jup}$) in a 1.7-day orbit about a G4 star. Our predicted transit depth variation caused by the atmosphere of WASP-122\,b suggests it is well suited to characterisation. WASP-123\,b is a hot-Jupiter ($0.90 M_{\rm Jup}$, $1.32 R_{\rm Jup}$) in a 3.0-day orbit around an old ($\sim 7$ Gyr) G5 star.
\end{abstract}


\section{Introduction}

The Wide Angle Search for Planets (WASP) survey is a prolific contributor to the field of exoplanet science having published the discovery of 104 planets to date. Our effective magnitude range of $9<$ V $< 13$ means that WASP systems are conducive to further study. Examples from the extremes of this range are the bright WASP-33 (V = 8.3; \citealt{2010MNRAS.407..507C-wasp33}) and WASP-18 (V = 9.3; \citealt{2009Natur.460.1098H-wasp18}) and the relatively dim WASP-112 ( V = 13.3; \citealt{2014arXiv1410.3449A-wasp112}). 

Here we present the discovery of: WASP-120\,b, a system with a star showing variable activity and a possibly eccentric planet orbit, WASP-122\,b, which offers a good opportunity for atmospheric study, and WASP-123\,b, which orbits an old star, $\sim 7$ Gyr.

\section{Observations}

\begin{figure}
 \centering
 \includegraphics[width=0.45\textwidth]{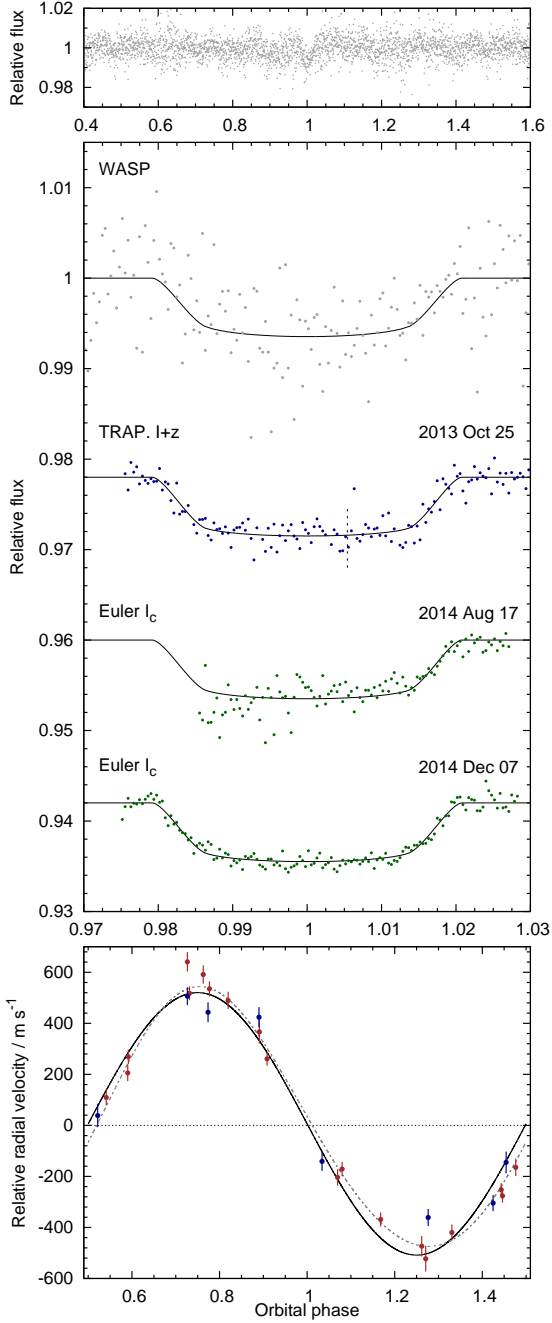}
 \caption{Discovery data for WASP-120\,b. {\it Top panel:} Phase folded WASP photometry for WASP-120. {\it Middle panel:} WASP discovery photometry (grey), TRAPPIST (blue) and EulerCam (green) follow-up photometry with our transit model over-plotted. The meridian flip in the TRAPPIST data has been corrected for and marked with a vertical dashed line. All photometric data have been binned with a duration of 2 minutes for clarity. {\it Bottom panel:} CORALIE radial velocity data from before (red circles) and after (blue triangles) the upgrade, over-plotted with the circular (black-solid) and eccentric (grey-dashed) solutions. }
\label{fig:wasp-120-graphs} 
\end{figure}

\begin{figure}
 \centering
 \includegraphics[width=0.45\textwidth]{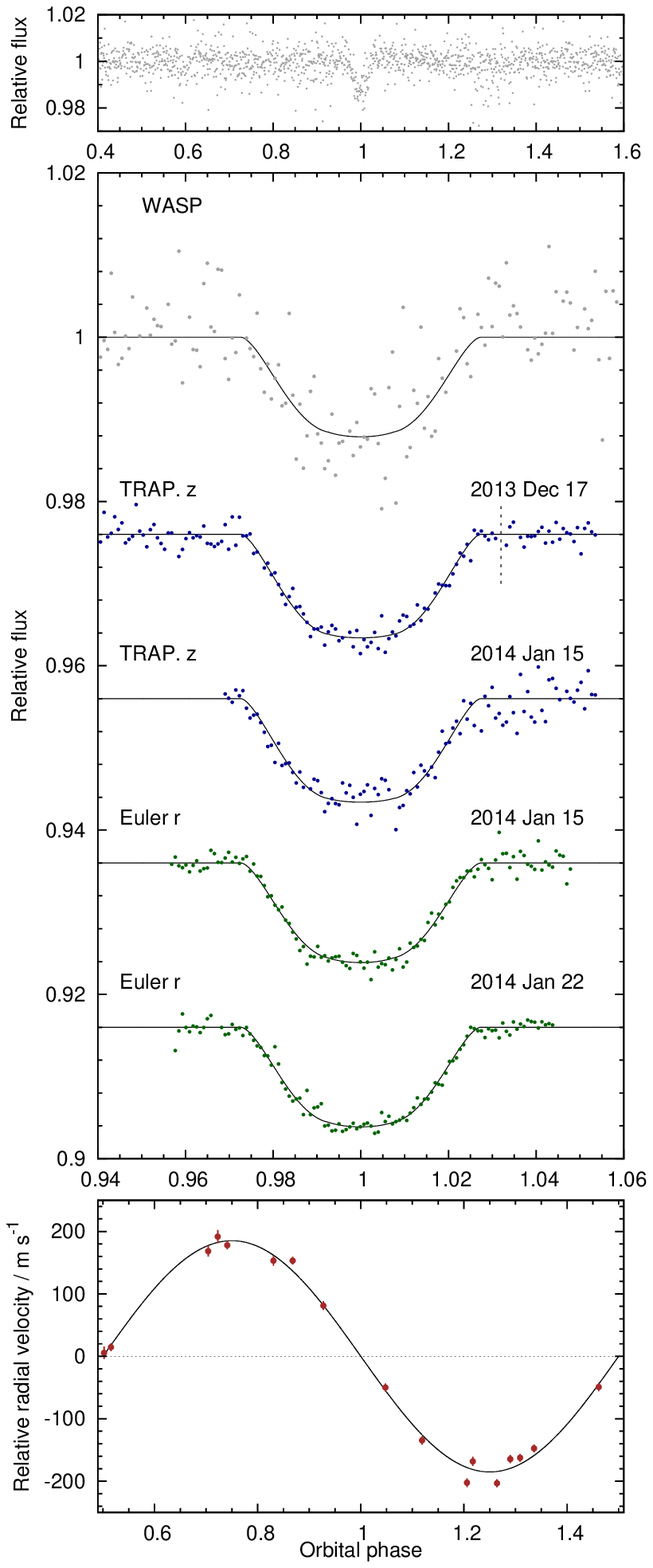}
 \caption{Discovery data for WASP-122\,b. {\it Top panel:} Phase folded WASP photometry for WASP-122. {\it Middle panel:} WASP discovery photometry (grey), TRAPPIST (blue) and EulerCam (green) follow-up photometry with our transit model over-plotted. The meridian flip in the TRAPPIST data has been corrected for and marked with a vertical dashed line. All photometric data have been binned with a duration of 2 minutes for clarity. {\it Bottom panel:} CORALIE radial velocity data, over-plotted with our circular solution.}
\label{fig:wasp-122-graphs} 
\end{figure}

\begin{figure}
 \centering
 \includegraphics[width=0.45\textwidth]{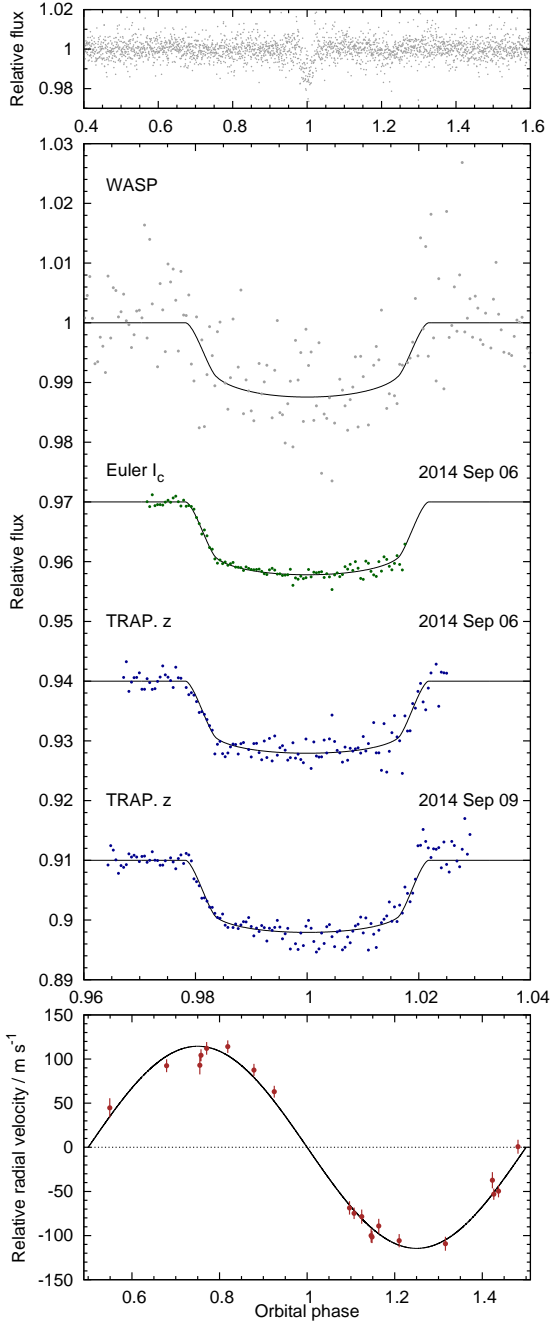}
\caption{Discovery data for WASP-123\,b. Caption as for Fig. \ref{fig:wasp-122-graphs}.}
\label{fig:wasp-123-graphs} 
\end{figure}

\begin{figure}
 \centering
 \includegraphics[width=0.45\textwidth]{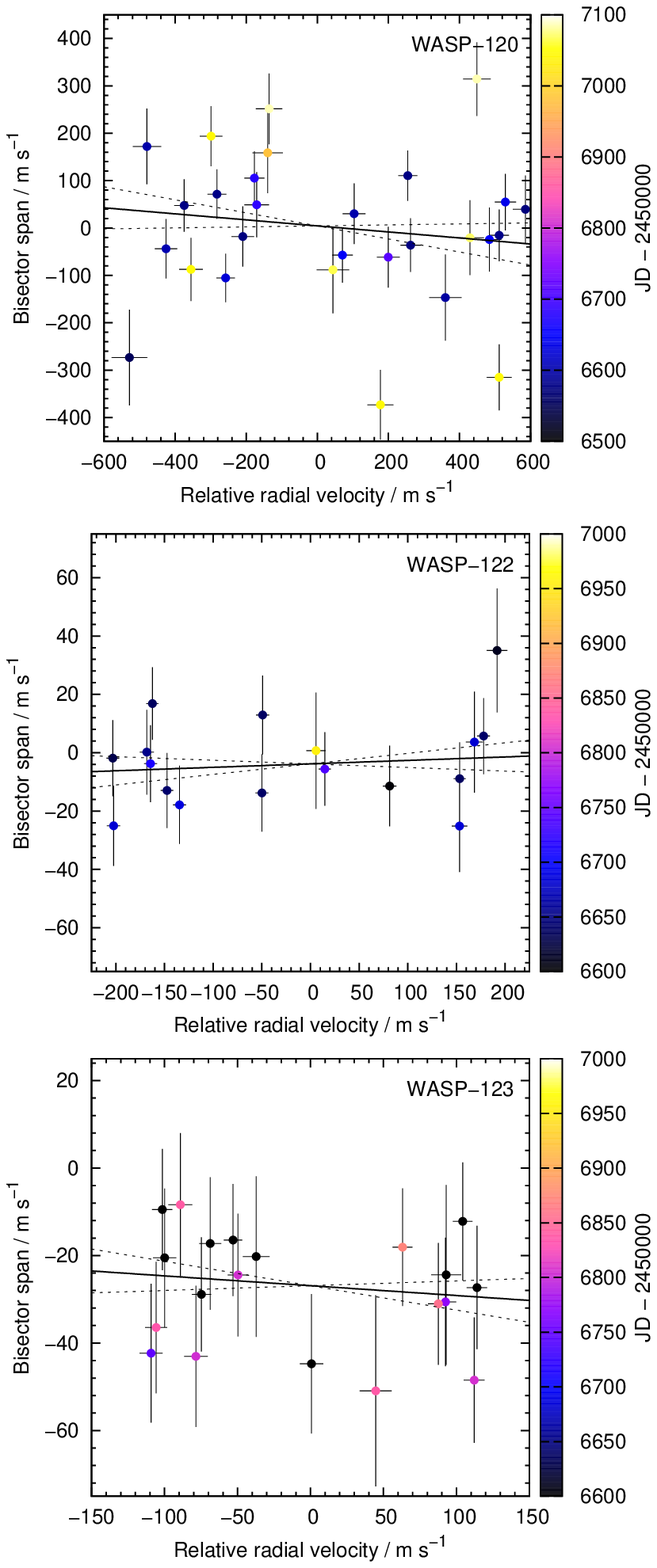} 
\caption{Bisector spans plotted against relative radial velocities for WASP-120 (top), WASP-122 (middle) and WASP-123 (bottom) showing no correlation. Solid lines are results of least-squares fits to the data, dashed lines are the $1\sigma$ uncertainties of the fits. Date of observation is denoted by point colour. The increased dispersion of points seen in more recent data for WASP-120 is attributed to an increase in stellar activity.}
\label{fig:bisector_spans}
\end{figure}

The transits of these planets were discovered in photometry gathered from the WASP-South installation hosted by the South African Astronomical Observatory. The WASP-South instrument is an array of 8 cameras using 200mm $f$/1.8 lenses to survey the sky at a cadence of $\sim 10$ minutes. For more information on the WASP instrument, see \cite{2006PASP..118.1407P}. The data were processed and searched for transits as described in \cite{2006MNRAS.373..799C} with candidate selection following the procedure in \cite{2007MNRAS.380.1230C}. Details of observations for each star in this paper can be found in Table \ref{tab:obs}. The phase-folded WASP data are displayed in the top panels of Figs. \ref{fig:wasp-120-graphs}, \ref{fig:wasp-122-graphs} and \ref{fig:wasp-123-graphs}. We used the method of \cite{2011PASP.123.547M} to search the WASP photometry for modulations caused by star spots. We detected no rotational modulation above 2\,mmag which suggests that the hosts are inactive.

\begin{table}
\small
\caption{Observations of WASP-120, WASP-122 and WASP-123}
\label{tab:obs}
\begin{tabular}{llrl}\tableline
Date  & Source & N.Obs. /  & Comment\\
 & & Filter & \\
\tableline 
\multicolumn{1}{l}{WASP-120} & & \\
2006 Aug--2012 Jan & WASP-South & \multicolumn{1}{r}{27\,079}  \\
2013 Sep--2015 Mar & CORALIE &  \multicolumn{1}{r}{29}  \\
2013 Nov 12 & TRAPPIST & I+z & Meridian flip \\
2014 Aug 17 & EulerCam & I$_{\rm c}$ &  \\
2014 Dec 07 & EulerCam & I$_{\rm c}$ &  \\ 
\tableline 
\multicolumn{1}{l}{WASP-122} & & \\
2011 Oct--2012 Mar & WASP-South & \multicolumn{1}{r}{4\,834}    \\
2013 Nov--2014 Oct & CORALIE &  \multicolumn{1}{r}{17}    \\
2013 Dec 17 & TRAPPIST & z & Meridian flip \\
2014 Jan 15 & TRAPPIST & z & Slight cloud\\
2014 Jan 15 & EulerCam & r & Slight cloud \\
2014 Jan 22 & EulerCam & r & Slight cloud \\
\tableline 
\multicolumn{1}{l}{WASP-123} & & \\
2006 May--2012 Jun & WASP-South & \multicolumn{1}{r}{13\,267}   \\
2013 Sep--2014 Aug & CORALIE &  \multicolumn{1}{r}{20}   \\
2014 Sep 06 & EulerCam & I$_{\rm c}$ &  \\
2014 Sep 06 & TRAPPIST & z & \\
2014 Sep 09 & TRAPPIST & z & \\
\tableline 
\end{tabular} 
\end{table}

We obtained spectra of the three stars with the CORALIE spectrograph on the 1.2-m Swiss telescope as outlined in Table \ref{tab:obs}. We used these data to measure radial velocity (RV) variations and confirm the planetary nature of the candidates (Table \ref{tab:follow_up_RV}; bottom panel of Figs. \ref{fig:wasp-120-graphs}, \ref{fig:wasp-122-graphs} and \ref{fig:wasp-123-graphs}). We obtained 9 of the WASP-120 spectra after the spectrograph was upgraded in November 2014. The lack of correlation between the bisector spans and RVs (Fig. \ref{fig:bisector_spans}) indicate that the RV variations are not a result of blended eclipsing binaries. For example, \cite{2002A&A...392..215S} found a brown dwarf mass companion that produces a correlation between the RVs and bisector spans with a gradient of 0.67. The largest gradient from amongst our planets is an order of magnitude smaller and not significant; $0.06 \pm 0.07$ for WASP-120. While we cannot strictly rule out the case of blended planet hosting stars we can eliminate brown dwarf blends and more massive objects. 

We acquired the follow-up photometry needed to accurately determine the system parameters from the 0.6-m TRAPPIST telescope \citep{2011EPJWC..1106002G} and EulerCam \citep{2012A&A...544A..72L} on the Swiss telescope at La Silla, Chile. 
The TRAPPIST telescope's equatorial mount requires a meridian flip when the target culminates during an observation. These occurred at BJD = 2456609.725 during the transit of WASP-120 on 2013 Nov 12 and at BJD = 2456644.758 during the transit of WASP-122 on 2013 Dec 17. We account for any offsets introduced by treating them as two separate datasets during our analysis. The photometric data are presented in Table \ref{tab:follow_up_phot}. This follow up revealed a star within 2.2'' of WASP-120 which is $4.35 \pm 0.02$ magnitudes fainter in the I band and $3.89 \pm 0.02$ magnitudes fainter in the z band. CORALIE's fibres are 2'' in diameter and the RVs were obtained in good seeing, so the star is sufficiently distant that it did not contaminate the observations and thus could not cause a false positive.

\begin{table}
\small
\caption{Radial velocity data from CORALIE.}
\label{tab:follow_up_RV}
\begin{tabular}{rrrrcc}\tableline
\multicolumn{1}{c}{HJD} & \multicolumn{1}{c}{RV} & \multicolumn{1}{c}{Error} & \multicolumn{1}{c}{BS}  & \multicolumn{1}{c}{Target} \\
\multicolumn{1}{c}{$-$ 2\,540\,000} & \multicolumn{1}{c}{(km s$^{-1}$)} & \multicolumn{1}{c}{(km s$^{-1}$)} & \multicolumn{1}{c}{(km s$^{-1}$)} & \multicolumn{1}{c}{Name} \\
\tableline \\
6552.902673 & 19.30305 & 0.05043 & $-$0.27331 & WASP-120 \\
6572.735422 & 20.41666 & 0.03569 & 0.03952 & WASP-120 \\
6573.843533 & 19.62205 & 0.03173 & $-$0.01785 & WASP-120 \\
... & ... & ... & ... & ... \\
... & ... & ... & ... & ... \\
6871.746191 & 16.99747 & 0.00673 & $-$0.01808 & WASP-123 \\
\tableline
\end{tabular} 
\tablecomments{Data available in this format at ADS. The data are provided \\ to the full precision used in our calculations but times are only accurate \\ to a few seconds at best.} 
\end{table}

\begin{table*}
\centering
\small
\caption{Follow-up photometry from TRAPPIST and EulerCam.}
\label{tab:follow_up_phot}
\tabcolsep=0.11cm
\begin{tabular}{rrrrrrrrrccc}\tableline
\multicolumn{1}{c}{HJD$_{\rm UTC}$}& \multicolumn{1}{c}{Norm.} & \multicolumn{1}{c}{Error} & \multicolumn{1}{c}{$\Delta$X}  & \multicolumn{1}{c}{$\Delta$Y} & \multicolumn{1}{c}{Airmass} & \multicolumn{1}{c}{Target} & \multicolumn{1}{c}{Sky Bkg.} & \multicolumn{1}{c}{Exp.} & \multicolumn{1}{c}{Target Name} & \multicolumn{1}{c}{Instrument} & \multicolumn{1}{c}{Band}\\
\multicolumn{1}{c}{$-$ 2\,540\,000}& \multicolumn{1}{c}{Flux} &  & \multicolumn{1}{c}{Position}  & \multicolumn{1}{c}{Position} &  & \multicolumn{1}{c}{FWHM} & \multicolumn{1}{c}{(Counts)} & \multicolumn{1}{c}{Time (s)} &  & & \\
\tableline \\
6887.720753 & 0.990347 & 0.002159 & $-$1.91 & $-$2.21 & 2.64 & 11.12 & 71.30 & 50.00 & WASP-120 & EulerCam & IC \\
6887.721534 & 0.996407 & 0.002145 & 0.25 & $-$1.40 & 2.62 & 12.18 & 70.00 & 50.00 & WASP-120 & EulerCam & IC \\
6887.722324 & 0.994713 & 0.002114 & 0.06 & $-$1.04 & 2.60 & 10.85 & 68.82 & 50.00 & WASP-120 & EulerCam & IC \\
... & ... & ... & ... & ... & ... & ... & ... & ... & ... & ... & ... \\
... & ... & ... & ... & ... & ... & ... & ... & ... & ... & ... & ... \\
6910.766640 & 0.992689 & 0.008096 & $-$0.92 & $-$0.64 & 2.91 & 4.08 & 293.86 & 13.00 & WASP-123 & TRAPPIST & z \\
\tableline
\end{tabular} 
\tablecomments{Data available in this format at ADS. The data are provided to the full precision used in our calculations but times are only accurate to a few seconds at best.}
\end{table*}

\section{Analysis}

\subsection{Stellar Parameters}

We determined the atmospheric parameters of each host star by analysing the co-added CORALIE spectra after correcting them for shifts due to the radial motion of the star using the measured RVs. Our spectral analysis followed procedures given in \cite{2013MNRAS.428.3164D}. For each star we obtained the effective temperature, $T_{\rm eff}$, using the H$\alpha$ line, $\log g$ from the Na D and Mg b lines and iron abundances from the analysis of equivalent width measurements of several unblended Fe~{\sc i} lines. We found the projected rotation velocity, $V \sin i$, by fitting the profiles of the Fe~{\sc i} lines after convolving with the instrumental resolution ($R$ = 55\,000) and a macroturbulent velocity adopted from the calibration of \cite{2014MNRAS.444.3592D}. 

\subsection{System Parameters}

We used a Markov chain Monte Carlo (MCMC) code to determine the system parameters using the discovery and follow-up photometry with RVs as described by \cite{ccam_etal2007} and \cite{2015A&A...575A..61A}.

\begin{table*}
 \centering
\small
\caption{Limb-darkening parameters extrapolated using the $T_{\rm LD}$ resulting from each analysis.}
\label{tab:LD-values}
\begin{tabular}{llllcccc}
\hline
Planet & Instrument & Instrument Band & Claret band & $a_{1}$ & $a_{2}$ & $a_{3}$& $a_{4}$  \\
\hline
WASP-120 & WASP & Broadband (400-700 nm) & Cousins R &  0.136 & 1.286 & $-$1.188 & 0.391 \\     
         & TRAPPIST & I+z & Sloan z & 0.221 & 0.827 & $-$0.760 & 0.226 \\     
         & EulerCam & Cousins I & Cousins I & 0.200 & 0.951 & $-$0.863 & 0.263 \\ \hline
WASP-122 & WASP & Broadband (400-700 nm) & Cousins R & 0.717 & $-$0.503 & 1.076 & $-$0.515 \\     
         & TRAPPIST & z & Sloan z & 0.799 & $-$0.743 & 1.095 & $-$0.492 \\     
         & EulerCam & Gunn R & Cousins R & 0.717 & $-$0.503 & 1.076 & $-$0.515\\ \hline      
WASP-123 & WASP & Broadband (400-700 nm) & Cousins R & 0.683 & $-$0.405 & 0.957 & $-$0.473 \\     
         & TRAPPIST & z & Sloan z & 0.766 & $-$0.664 & 1.010 & $-$0.462 \\     
         & EulerCam & Cousins I & Cousins I & 0.763 & $-$0.639 & 1.059 & $-$0.491 \\ \hline      
 
\end{tabular}

\end{table*}

For each system we modelled our transit lightcurves using the formulation of \cite{mandel+agol2002} and accounted for limb-darkening using the four-parameter non-linear law of \cite{Claret2000,Claret2004}. The photometric bands and limb-darkening coefficients used in the lightcurve models are detailed in Table \ref{tab:LD-values}. 

We used {\sc bagemass} \citep{2015A&A...575A..36M} to compare $\rho_{s}$, determined from the transit lightcurves coupled with the spectroscopic values of [Fe/H] and $T_{\rm eff}$, to stellar models in order to estimate the mass of the star.  {\sc bagemass} also gives an estimate of the age of the system.

To calculate the distance we use the apparent K$_{\rm s}$ band magnitude from \citet{2006AJ....131.1163S}, the radius of the star from Table~\ref{tab:Spect} and the angular diameter of the star based on the calibration of K-band surface brightness -- effective temperature relation from \citet{2004A&A...426..297K}. We assume that interstellar reddening is negligible and that K$ = {\rm K}_{\rm s} +0.044$.

\begin{figure*}
 \centering
 \includegraphics[width=0.98\textwidth]{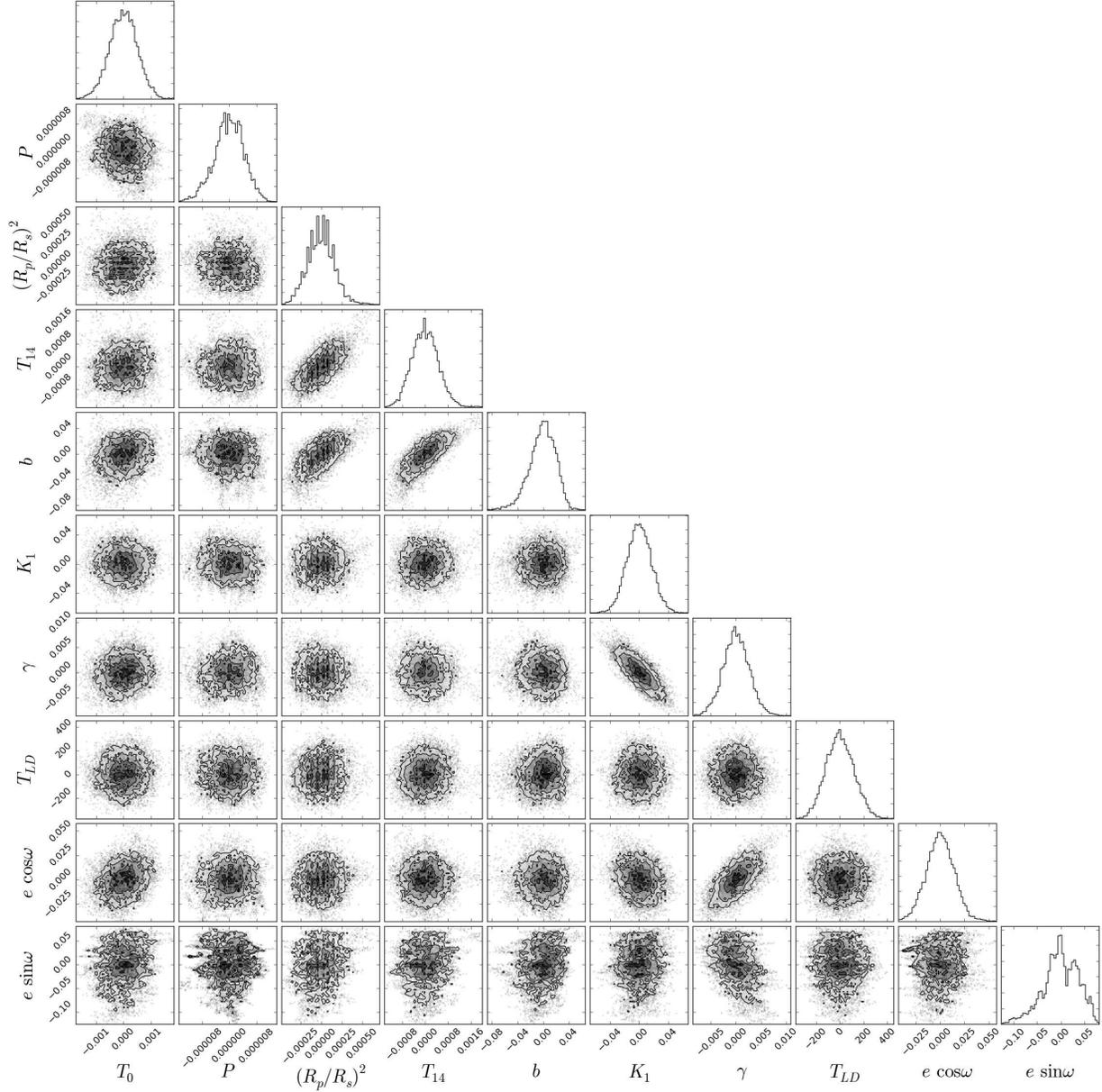}
\caption{Plots showing the relationships of various jump parameters $T_{0}$, $P$, $(R_{P}/R_{s})^{2}$, $T_{14}$, $b$, $K_{1}$, $\gamma$,  ${[}\textrm{Fe/H}{]}$, $T_{\rm LD}$, $\sqrt{e}\cos \omega$ and $\sqrt{e} \sin \omega$ from the analysis of WASP-120. All distributions have had the mean value from Table \ref{tab:Spect} subtracted. Plots were prepared with a modifed version of trangle.py by \protect\cite{dan_foreman_mackey_2014_11020}}
 \label{fig:w120-corner}
\end{figure*}

\begin{figure*}
 \centering
 \includegraphics[width=0.98\textwidth]{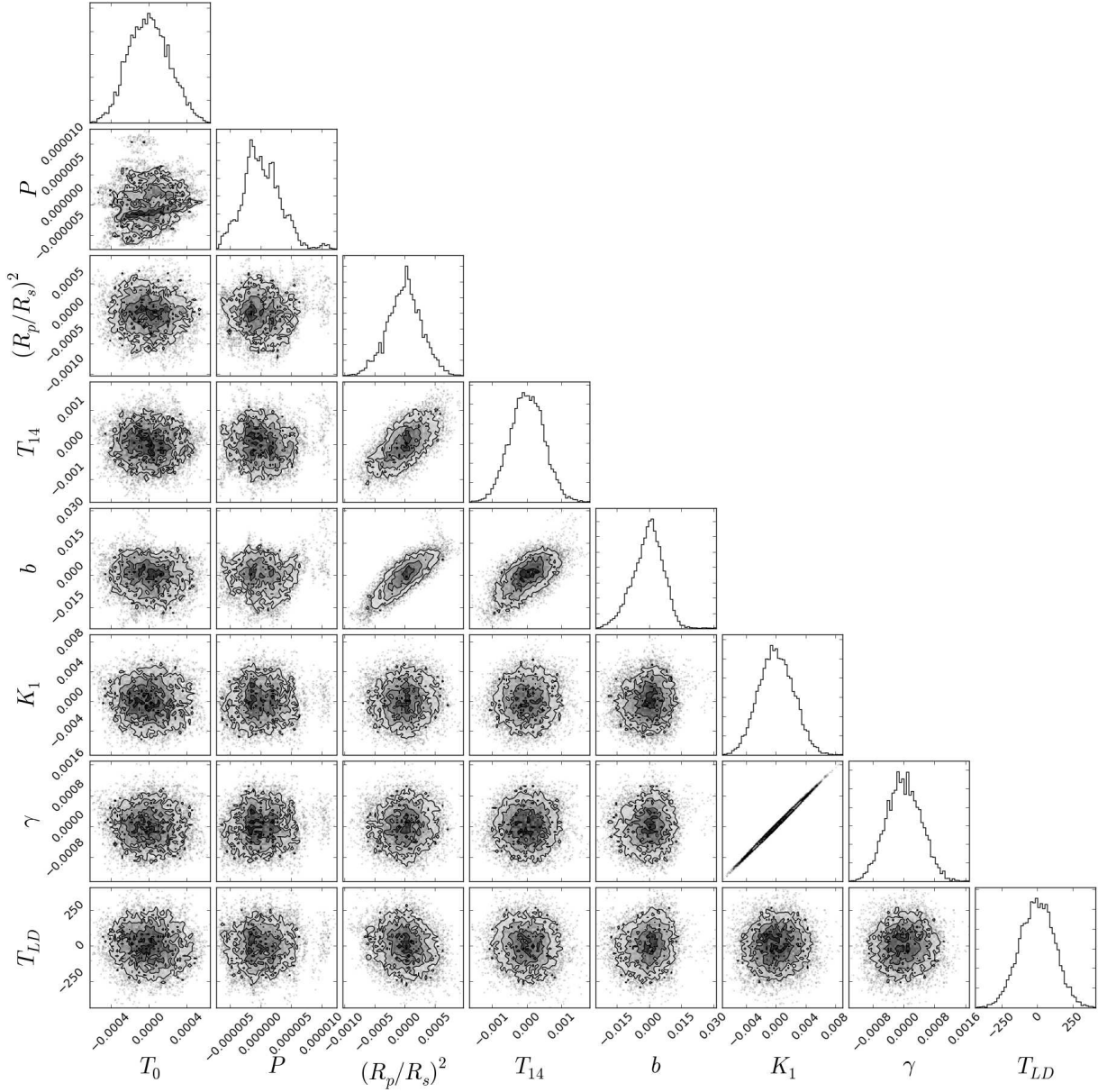}
\caption{Caption as for Fig. \ref{fig:w120-corner}. As we enforced a circular orbit we did not fit $\sqrt{e}\cos \omega$ and $\sqrt{e} \sin \omega$.}
 \label{fig:w122-corner}
\end{figure*}

\begin{figure*}
 \centering
 \includegraphics[width=0.98\textwidth]{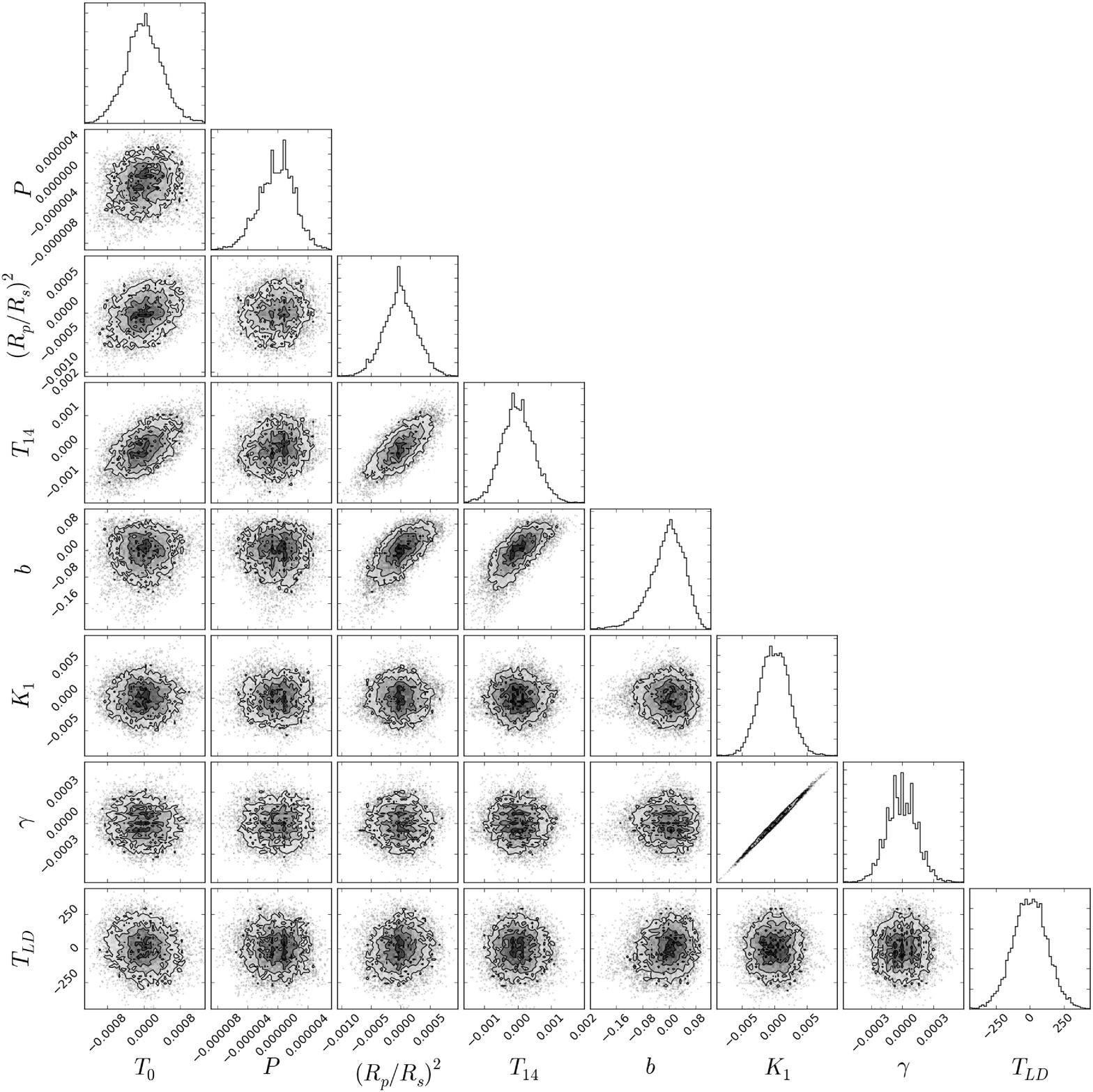}
\caption{Caption as for Fig. \ref{fig:w122-corner}}
 \label{fig:w123-corner}
\end{figure*}

The free parameters in our MCMC analysis were $T_{0}$, $P$, $(R_{P}/R_{s})^{2}$, $T_{14}$, $b$, $K_{1}$, $\gamma$,  ${[}\textrm{Fe/H}{]}$ and $T_{\rm LD}$. Here $T_{0}$ is the epoch of mid-transit, $P$, is the orbital period, $(R_{P}/R_{s})^{2}$ is the planet-to-star area ratio, $T_{14}$ is the total transit duration, $b$ is the impact parameter of the planet's path across the stellar disc, $K_{1}$ is the reflex velocity semi-amplitude, $\gamma$ is the systemic velocity, ${[}\textrm{Fe/H}{]}$ is the stellar metallicity and $T_{\rm LD}$ is the limb-darkening temperature. $T_{\rm LD}$ and ${[}\textrm{Fe/H}{]}$ were constrained by the spectroscopic values of $T_{\rm eff}$ and ${[}\textrm{Fe/H}{]}$. $T_{\rm LD}$ was used by the MCMC to interpolate limb-darkening coefficients at each step from the Claret limb-darkening tables for the appropriate photometric band of each lightcurve. At each step of our MCMC these values were perturbed by a small random value and the $\chi^{2}$ of the model based on the new values was calculated. If this lead to a lower $\chi^{2}$ the step was accepted while a larger value would be accepted with a probability proportional to $\exp(-\Delta \chi^{2}/2)$. Our final values were calculated from the medians of the posterior distributions with uncertainties corresponding to the $1 \sigma$ confidence intervals. When we allow the MCMC to explore eccentric solutions we fit $\sqrt{e}\cos \omega$ and $\sqrt{e} \sin \omega$ to ensure a uniform probability distribution. Our results for each system are in the lower part of Table \ref{tab:Spect} and corner plots of the jump parameter posterior distributions of each analysis in Figures \ref{fig:w120-corner}, \ref{fig:w122-corner} and \ref{fig:w123-corner}.

We checked for trends in the derived transit depths with respect to the colour of the observational band for each star by running each lightcurve through our MCMC separately. The depths for WASP-122 and WASP-123 agree to within 1$\sigma$ of the depth derived from the combined analysis. The depths from this analysis of the two complete lightcurves of WASP-120 show a slight difference of $(1.2 \pm 0.4 )\times 10^{-3}$ which could be accounted for by a low level of inherent stellar variability in either the host star or faint, nearby companion.

\section{WASP-120} 

WASP-120\,b is a 4.85-$M_{\rm Jup}$, 1.73-$R_{\rm Jup}$ planet orbiting a moderately bright ($V = 11.0$) F5 star. The effective temperature of WASP-120 places it in the lithium gap \citep{2004AJ.128.2435B}, so we cannot estimate the age of this star based on the lithium abundance. Using the star's Tycho B$-$V colour and rotation period from its $V\sin i$ and radius from our MCMC we use gyrochronology calibration of \cite{2007ApJ...669.1167B} to estimate an age of $0.7 \pm 0.6$ Gyr. For comparison, the calibration of \cite{2008ApJ...687.1264M} gives $1.0 \pm 1.8$ Gyr. We cannot apply the calibration of \cite{2009ApJ...695..679M} as the star's colour results in a term requiring the logarithm of a negative value. Using {\sc bagemass} we find an age of $2.6 \pm 0.5$ Gyr which is consistent with that of the Mamajek \& Hillenbrand calibration. 

\begin{figure*}
 \centering
 \includegraphics[width=0.98\textwidth]{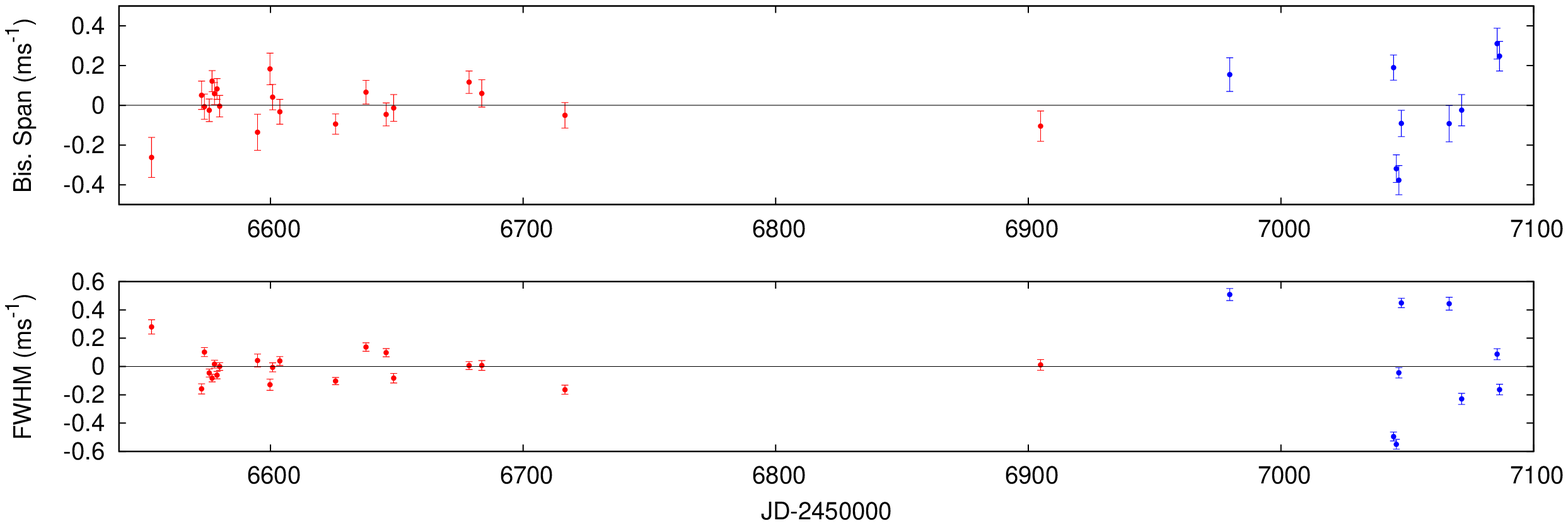}
\caption{Full width at half-maximum (FWHM) and bisector spans of spectra as a function of time for WASP-120. The increase in scatter in more recent data is taken as an indication the star may be entering a phase of increased activity. Red circles are data taken before the CORALIE upgrade and blue triangles are those taken after, the black dotted lines denote the date CORALIE was back on sky after the upgrade. The data have all had the mean of their distribution subtracted before plotting.}
 \label{fig:activity}
\end{figure*}

The  FWHM of the lines in the spectra and the  bisector spans show more scatter in the later data, after the CORALIE upgrade (Fig. \ref{fig:activity}), suggesting that the star may have become more active, and therefore have variable activity like the Sun. It is unlikely that the increased scatter is caused by the change to CORALIE since datasets on other stars don't show an increased scatter (e.g. recent RV data taken of WASP-47; \citealt{2016A&A...586A..93N}).

Due to the upgrade to CORALIE we partitioned the data into two sets. We added jitter of $5.1 \pm 0.2$ m\,s$^{-1}$ to the older data and $6.3 \pm 0.8$ m\,s$^{-1}$ to the newer data. These values were adopted such that both datasets gave reduced $\chi^{2}$ values of one compared to a circular-orbit solution, and are in keeping with jitter determined for similar stars by \cite{2005PASP..117..657W}. The uncertainties were estimated using a jackknife resampling. The difference between the jitter values of $0.8 \pm 0.8$ is consistent with zero, thus the jitter values are consistent to 1$\sigma$.

Our resulting orbital solution had an eccentricity of $0.057^{+0.022}_{-0.018}$. This is significantly non-zero at 3.3$\sigma$, while a Lucy-Sweeney test \citep{1971AJ.....76..544L} gives a probability of only 0.1\% that the orbit is circular. We note, though, that this result is somewhat dependent on the jitter values used. Fitting with no jitter increased the eccentricity to $0.068 \pm 0.014$. This led to a higher apparent significance ($4.9\sigma$) though the Lucy-Sweeney test no longer excluded the circular solution since neither solution is a good fit to the data. We thus adopt the value of eccentricity with jitter added but regard this as needing confirmation. 

\begin{figure}
 \centering
 \includegraphics[width=0.48\textwidth]{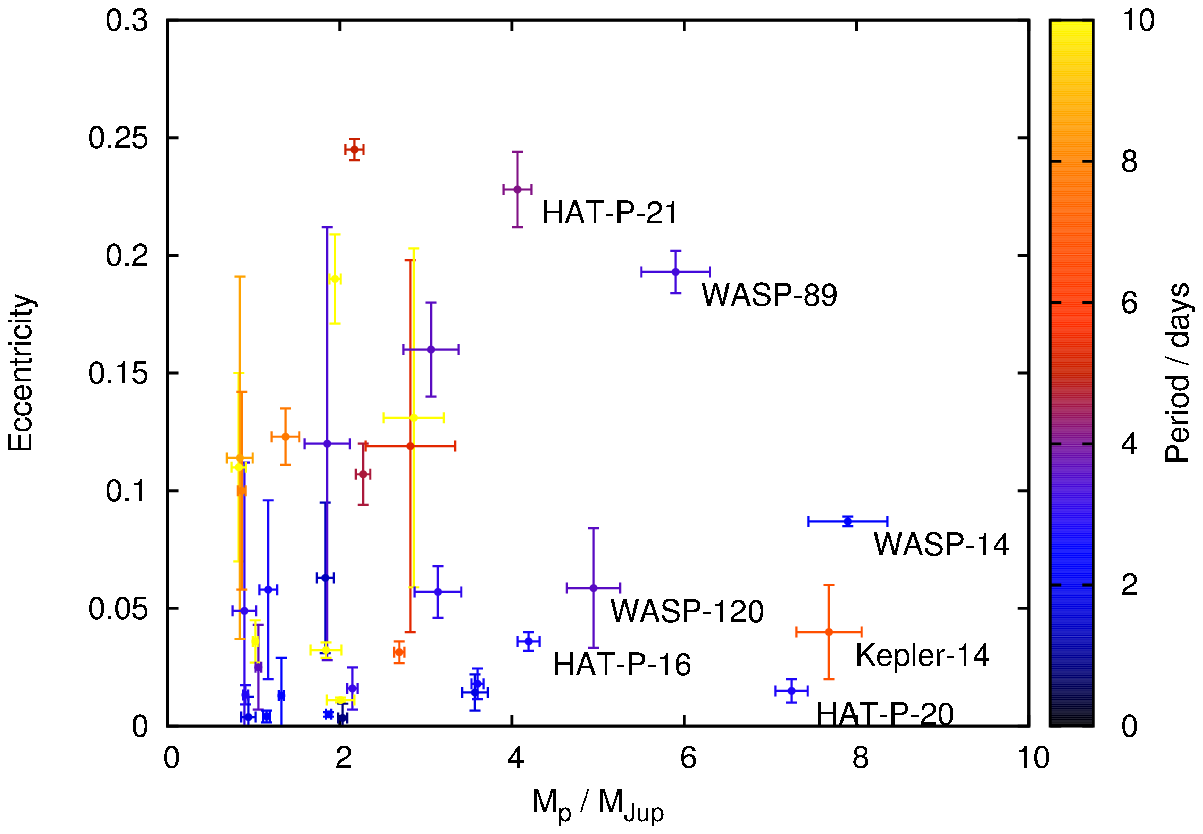}
\caption{Masses, eccentricities and periods of transiting planets with non-zero eccentricities quoted in literature and masses $> 0.5M_{{\rm Jup}}$ mass. The most convincing eccentricities are those associated with more massive planets. Notable examples and WASP-120\,b  are labelled. Data from TEPCat \citep{2011MNRAS.417.2166S}.}
 \label{fig:mass_p_ecc}
\end{figure}

More massive planets, such as WASP-120\,b at 5.0-$M_{\rm Jup}$, often have eccentric orbits (Fig \ref{fig:mass_p_ecc}), though it is unclear if the correlation is due to a real phenomenon or observation bias as the eccentricities of more massive planets are easier to detect from RVs. A conclusive determination of this system's eccentricity would come from observation of an occultation which we expect to be delayed by $2.6 \pm 1.1$ hours if our eccentricity value is accurate. Based on the equilibrium temperature of the planet, we estimate occultation depths in the {\it Spitzer} 3.6  $\mu m$ and 4.5 $\mu m$ bands of approximately 780 and 960 ppm, respectively. Recent observations with {\it Spitzer} \citep{2014ApJ...796..115Z,2015ApJ...805..132D} show detecting such an occultation is easily achievable.

\begin{figure*}
 \centering
 \includegraphics[width=0.98\textwidth]{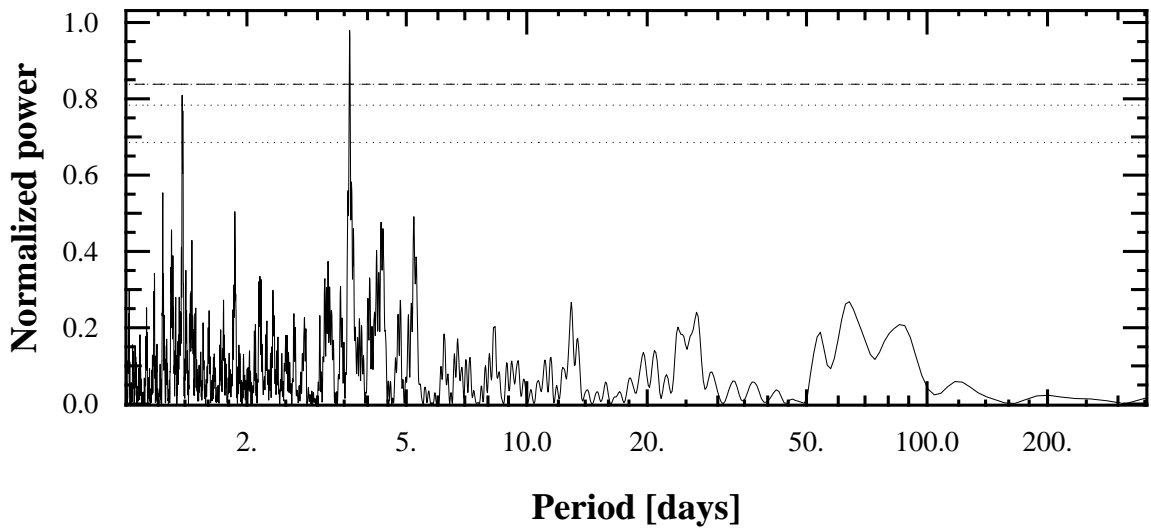}
\caption{Periodogram for the RV data of the WASP-120 system. The only significant peak is that corresponding to the planetary period.}
 \label{fig:ls-w120}
\end{figure*}

The faint star close to WASP-120 may provide evidence in support of the planet having undergone high-eccentricity migration due to Kozai-Lindov cycles. According to \cite{2008ApJS..178...89D}, the companion's colour is consistent with a 0.6 $M_{\odot}$, K9 star at the same distance as WASP-120. Assuming this to be the case, the on sky separation gives a minimum separation of 950 AU. Using the equations of \cite{2007ApJ...669.1298F}, a star of this mass could induce Kozai-Lindov cycles if the planet's original orbital distance was 14.5 AU or greater. As the periodogram of the RVs for WASP-120 show no significant peaks beyond that of the planet (Fig. \ref{fig:ls-w120}) we checked for the presence of an additional, long period, object in the system by fitting a linear trend to the residuals of the RVs. The result was an RV drift, $\dot{\gamma}$, of $(84 \pm 73)$ m\,s$^{-1}$\,yr$^{-1}$ which is consistent with zero at $\sim 1.2\sigma$. Following \cite{2014ApJ...781...28M}, for a planet on a circluar orbit:

\small
\begin{equation}
\dot{\gamma} = (6.57 {\rm m\,s^{-1}\,yr^{-1}}) \left(\frac{M_{2}}{M_{\textrm{Jup}}}\right) \left(\frac{a_{2}}{5 {\rm AU}}\right)^{-2} \sin i_{2}
\end{equation}
\normalsize

Therefore, if there is an additional object in the system it has a mass, $M_{2}$, semi-major axis $a_{2}$, and inclination, $i_{2}$, such that, $M_{2} \sin i_{2} / a_{2}^{2} \lesssim 0.51 M_{\rm Jup}$ AU$^{-2}$.

Notable examples of massive planets with confidently detected eccentricities are HAT-P-16\,b \citep{2010ApJ...720.1118B}, HAT-P-21\,b \citep{2011ApJ...742..116B}, WASP-14\,b \citep{2009MNRAS.392.1532J} and WASP-89\,b \citep{2015AJ....150...18H}. All of these are in sub 7-day orbits with masses $> 4 M_{\rm Jup}$. Also notable are HAT-P-20\,b \citep{2011ApJ...742..116B}, which has the smallest eccentricity of the group and Kepler-14\,b with the longest orbital period at 6.79-days \citep{2011ApJS..197....3B}. Like WASP-120, 3 of these 6 systems are known to have other stars nearby; HAT-P-16, WASP14 and Kepler-14 \citep{2015A&A...575A..23W,2015A&A...579A.129W,2015ApJ...800..138N,2011ApJS..197....3B}. However, this sample is too small to draw conclusions about a link between orbital eccentricity and the presence of a stellar-mass neighbour.

It is thought that stars with effective temperatures cooler than 6200 K have convective envelopes which enhance orbit circularisation/re-alignment \cite{2010ApJ...718L.145W}. This could erase any correlation between the type of orbit a planet is in and the presence of a further companion. 

In a study on the prevalence of multiple stars in planetary systems \cite{2015ApJ...800..138N} found that, of their sample of hot host stars ($T_{\rm eff} > 6200 K$) with evidence of misaligned or eccentric planet orbits $59\% \pm 17\%$ had companions while $83\% \pm 14\%$ of their well-aligned/circular orbit sample had companions. When \cite{2015ApJ...800..138N} considered just the spin-orbit alignment of these systems, $73\% \pm 15\%$ of misaligned systems had companions as opposed to $53\% \pm 14\%$ of well aligned systems. They concluded that there is no evidence for a link between multiplicity and orbital eccentricity/misalignment, though so far the sample is just 18 hot stars, fewer when just those systems with measured spin-orbit alignments are used. 

The typical time-scale of orbital circularisation is expected to be shorter than that of tidal realignment meaning that observations of spin-orbit misalignment may provide a better record of migration pathway than eccentricity. Therefore, increasing the number of hot-host star systems with measured spin-orbit alignment that have been evaluated for stellar multiplicity could change the current picture. 

\section{WASP-122}

WASP-122\,b is a $1.28 M_{\rm Jup}$, $1.74 R_{\rm Jup}$ planet orbiting a moderately bright (V = 11.0), metal rich ([Fe/H] = $+0.32 \pm 0.09$), G4 star. WASP-122 is depleted in lithium ($\log A (Li) <1.0$) and so must be several Gyr old \citep{2005A&A.442.615S}. Using the star's colour and rotational period from its $V\sin i$, we calculate a gyrochronological ages of $2.3 \pm 1.4$ Gyr \citep{2007ApJ...669.1167B}, $2.8 \pm1.4$ \citep{2008ApJ...687.1264M} and $2.9 \pm 4.8$ \citep{2009ApJ...695..679M}.  
Using {\sc bagemass} we find two possible solutions. Approximately 75\% of the Markov chain output by {\sc bagemass} favour a mass of $1.24 \pm 0.04$ $M_{\odot}$ and an age of $5.11 \pm 0.80$ Gyr. The other 25\% of the output prefer a solution giving a mass of $1.10 \pm 0.03$ $M_{\odot}$ and an age of $8.67 \pm 1.05$ Gyr. The favoured, younger, higher-mass solution is a better match to, though still older than, the gyrochronological ages.  

WASP-122\,b presents a good target for atmospheric characterisation via transmission spectroscopy. Assuming the atmosphere is isothermal and adequately described as an ideal gas we can calculate the atmospheric scale height, $H$, using: 

\begin{equation}
 H = \frac{k T_{\rm eq}}{g \mu} 
\end{equation}

Here, $k$ is Boltzmann's constant, $T_{\rm eq}$ is the planetary equilibrium temperature, $g$ is the planetary surface gravity and $\mu$ is the mean molecular mass of the atmosphere. We can use $H$ to predict the transit depth variation due the addition of  one atmospheric scale height to the planetary radius. In the case of WASP-122\,b this is 142 ppm. The same calculation for the well-studied HD209458\,b yields a variation of $\approx 200$ ppm. \cite{2013ApJ...774...95D} found evidence for water absorption on this scale in HD209458\,b. While WASP-122 is dim by comparison to HD209458 constraints have been put on the atmospheric compositions of planets with similarly bright hosts. For example, studies of WASP-12 (V = 11.6, ) show evidence of aerosols and a lack of TiO \citep{2013MNRAS.436.2956S} as well as placing constraints on the C/O ratio of the planet \citep{2015ApJ...814...66K} which has been suggested may be an indicator of formation environment. We predict occultation depths in 3.6 $\mu m$ and 4.5 $\mu m$ {\it Spitzer} bands of 2100 and 2500 ppm respectively. In the K-band we predict a depth of 1000 ppm. Similar K-band depths have been detected, for example that of WASP-10b \citep{2015A&A...574A.103C}, making ground-based follow up possible. Such observations of WASP-122\,b stand to shed light on our understanding of atmospheric albedo and opacity sources as well as its formation history.

\section{WASP-123} 
       
WASP-123\,b is a $0.90$-$M_{\rm Jup}$, $1.32$-$R_{\rm Jup}$ planet orbiting a moderately bright (V = 11.1), G5 star with a super-solar metal abundance ([Fe/H] = $+0.18 \pm 0.08$). WASP-123 is depleted in lithium ($\log A (Li) <0.5$) suggesting an age of several Gyr. This star falls into an area of parameter space for which gyrochronology is poorly calibrated \citep{2014EAS....65..289J}. The Barnes calibration gives an age greater than the present age of the universe and Mamajek and Meibom calibrations are not applicable as the star's colour results in the calibrations requiring the logarithm of a negative value. The age we derive using {\sc bagemass}, $6.9 \pm 1.4$ Gyr, supports an advanced age. Planets of similar mass and radius are not uncommon and are frequently found in orbits $\sim 3$-days around such host stars. This makes WASP-123 a typical example of a hot-Jupiter system. However, even these can prove surprising (e.g. WASP-47; \citealt{2015ApJ...812L..18B,2016A&A...586A..93N}) and/or contribute as vital controls to other studies.

\acknowledgements

WASP-South is hosted by the South African Astronomical Observatory and we are grateful for their ongoing support and assistance. Funding for WASP comes from consortium universities and from the UK's Science and Technology Facilities Council. O.D.T is also funded by the UK's Science and Technology Facilities Council. TRAPPIST is funded by the Belgian Fund for Scientific Research (Fond National de la Recherche Scientifique, FNRS) under the grant FRFC 2.5.594.09.F, with the participation of the Swiss National Science Foundation (SNF). M.G. and E.J. are FNRS Research Associates. A.H.M.J.T. is a Swiss National Science Foundation Fellow under grant P300P2-147773. L.D. acknowledges the support of the F. R. I. A. fund of the FNRS. The Swiss {\it Euler} Telescope is operated by the University of Geneva, and is funded by the Swiss National Science Foundation. M.L acknowledges support of the European Research Council through the European Union's Seventh Framework Programme (FP7/2007-2013)/ERC grant agreement number 336480.

\begin{table*}
\small
\caption{Stellar and planetary parameters parameters determined from spectra and MCMC analysis. Spectral parameters have formal uncertainties while parameters found via MCMC are the median values of the posterior distributions with an uncertainty corresponding to the $1\sigma$ confidence interval.}
\label{tab:Spect}
\begin{tabular}{lccc}\tableline
Spectroscopic Parameter & WASP-120 &  WASP-122 & WASP-123\\
\tableline
Tycho-2 ID & 8068-01208-1 & 7638-00981-1 & 7427-00581-1\\
USNO-B ID & 0441-0033568 & 0475-0113097 & 0571-1147509\\
RA (J2000) & 04:10:27.85 & 07:13:12.34 & 19:17:55.04 \\
Dec (J2000) & $-$45:53:53.5  & $-$42:24:35.1 & $-$32:51:35.8 \\
V Magnitude & 11.0 & 11.0 & 11.1 \\
Tycho (B$-$V) colour  & $ 0.523 \pm 0.083 $ & $ 0.78 \pm 0.11 $ & $ 0.48 \pm 0.17 $ \\
Spectral Type & F5 & G4 & G5\\
Distance (pc) & $437 \pm 21$ & $266 \pm 10$ & $214 \pm 11$ \\
{\sc bagemass} Age (Gyr) & $2.6 \pm 0.5$ & $5.11 \pm 0.80$ & $6.9 \pm 1.4$ \\
Stellar Effective Temperature, $T_{\rm eff}$ (K) & $6450 \pm 120$ & $5720 \pm 130$ & $5740 \pm 130$\\
Stellar Surface Gravity, $\log {\rm g_{s}}$ & $4.3 \pm 0.1$ & $4.3 \pm 0.1$ & $4.3 \pm 0.1$ \\
Stellar Metallicity, ${[}\textrm{Fe/H}{]}$ & $-0.05 \pm 0.07$ &  $0.32 \pm 0.09$ & $0.18 \pm 0.08$ \\
Projected Rot. Vel., $V\sin i$  $(\textrm{km s}^{-1})$  &   $15.1 \pm 1.2$ &  $3.3 \pm 0.8$ & $1.0 \pm 0.7$ \\
Stellar Lithium Abundance, $\log A ({\rm Li})$ & $<1.2$ &  $<1.0$ & $<0.5$ \\
Micro turbulence ($\textrm{km s}^{-1}$) &  $1.5 \pm 0.1$ & $0.9 \pm 0.1$ & $1.0 \pm 0.1$ \\
Macro turbulence ($\textrm{km s}^{-1}$)&   $6.0 \pm 0.8$ & $3.4 \pm 0.5$ & $3.4 \pm 0.5$ \\
\tableline \tableline
MCMC Parameter & WASP-120 & WASP-122 & WASP-123\\
\tableline
\\Period, P (d)  & $3.6112706 \pm 0.0000043$ & $1.7100566^{+0.0000032}_{-0.0000026}$ & $2.9776412 \pm 0.0000023$ \\ 
Transit Epoch, $T_{0}$  & $6779.43556 \pm 0.00051$ & $6665.22401 \pm 0.00021$ & $6845.17082 \pm 0.00039$ \\ 
Transit Duration, $T_{14}$ (d) & $0.1483 \pm 0.0016$ & $0.09117 \pm 0.00082$ & $0.1289 \pm 0.0014$ \\ 
Scaled Semi-major Axis, $a/R_{s}$  & $5.90 \pm 0.33$ & $4.248 \pm 0.072$ & $7.13 \pm 0.25$ \\ 
Transit Depth, $(R_{p}/R_{s})^{2}$  & $0.00655 \pm 0.00016$ & $0.01386 \pm 0.00029$ & $0.01110 \pm 0.00027$ \\ 
Impact Parameter, $b$  &  $0.78 \pm 0.02$ & $0.8622 \pm 0.0071$ & $0.530 \pm 0.049$ \\ 
Orbital Inclination, $i$ (\degr)  & $82.54 \pm 0.78$ & $78.3 \pm 0.3$ & $85.74 \pm 0.55$ \\ 
Eccentricity, $e$  & $0.057^{+0.022}_{-0.018}$  & 0 (adopted;$< 0.08$ at $2\sigma$)  & 0 (adopted;$< 0.12$ at $2\sigma$)  \\
Argument of Periastron, $\omega$ (\degr) & $-27^{+48}_{-28}$& $-$ & $-$ \\
Systemic Velocity, $\gamma$ $(\rm kms^{-1})$ & $19.836 \pm 0.013 $	& $34.5934 \pm 0.0017 $	& $16.9344 \pm 0.0017$	\\
Semi-amplitude, $K_1$ $(\rm ms^{-1})$  & $509 \pm 17$ & $185.1 \pm 2.3$ & $114.2 \pm 2.2$ \\ 
Semi-major Axis, $a$ (AU)  & $0.0514 \pm 0.0007$ & $0.03005 \pm 0.00031$ & $0.04263 \pm 0.00074$ \\ 
Stellar Mass, $M_{s}$ $(M_{\odot})$  & $1.393 \pm 0.057$ & $1.239 \pm 0.039$ & $1.166 \pm 0.061$ \\ 
Stellar Radius, $R_{s}$ $(R_{\odot})$ & $1.87 \pm 0.11$ & $1.52 \pm 0.03$ & $1.285 \pm 0.051$ \\ 
Stellar Density, $\rho_{s}$ ($\rho_{\odot}$)  & $0.212^{+0.041}_{-0.031}$ & $0.351 \pm 0.018$ & $0.548 \pm 0.059$ \\ 
Stellar Surface Gravity, $\log(g_{s})$ (cgs) & $4.035 \pm 0.049$ & $4.166 \pm 0.016$ & $4.286 \pm 0.032$ \\ 
Limb-Darkening Temperature, $T_{\textrm{LD}}$ (K)  & $6440 \pm 120$ & $5750 \pm 120$ & $5740 \pm 130$ \\ 
Planet Mass, $M_{p}$ $(M_{\textrm{Jup}})$  & $4.85 \pm 0.21$ & $1.284 \pm 0.032$ & $0.899 \pm 0.036$ \\ 
Planet Radius, $R_{p}$ $(R_{\textrm{Jup}})$  & $1.473 \pm 0.096$ & $1.743 \pm 0.047$ & $1.318 \pm 0.065$ \\ 
Planet Density, $\rho_{p}$ ($\rho_{\textrm{Jup}})$  & $1.51^{+0.33}_{-0.26}$ & $0.243 \pm 0.019$ & $0.393 \pm 0.056$ \\ 
Planet Surface Gravity, $\log(g_{p})$ (cgs)  & $3.707 \pm 0.056$ & $2.985 \pm 0.022$ & $3.07 \pm 0.04$ \\ 
Planet Equilibrium Temperature, $T_{\textrm{eq}}$ (K) & $1880 \pm 70$ & $1970 \pm 50$ & $1520 \pm 50$ \\ 
\tableline 
\end{tabular} 
\end{table*}

\bibliography{disc-bibliography}{}
\bibliographystyle{apj}

\end{document}